\documentclass[fleqn,usenatbib]{mnras}

\usepackage{newtxtext,newtxmath} 

\usepackage[T1]{fontenc}
\usepackage{amsmath}

\DeclareRobustCommand{\VAN}[3]{#2}
\let\VANthebibliography\thebibliography
\def\thebibliography{\DeclareRobustCommand{\VAN}[3]{##3}\VANthebibliography}

\usepackage{graphicx}	
\usepackage{amsmath}	
\usepackage{gensymb}
\usepackage{booktabs}

\title[A transient ultraviolet outflow in UW CrB]{A transient ultraviolet outflow in the short-period X-ray binary UW CrB}

\author[Fijma et al.]{S. Fijma,$^{1}$\thanks{E-mail: s.c.fijma@uva.nl}
N. Castro Segura,$^{2}$
N. Degenaar,$^{1}$
C. Knigge,$^{2}$
N. Higginbottom,$^{2}$
J.~V. Hern\'{a}ndez Santisteban,$^{3}$
\newauthor
T.J. Maccarone$^{4}$
\\
$^{1}$Anton Pannekoek Institute for Astronomy, University of Amsterdam, Science Park 904, 1098 XH, Amsterdam, the Netherlands\\
$^{2}$Department of Physics \& Astronomy. University of Southampton, Southampton SO17 1BJ, UK\\
$^{3}$SUPA Physics and Astronomy, University of St. Andrews, North Haugh, KY16 9SS, UK\\
$^{4}$Department of Physics \& Astronomy, Texas Tech University, Box 41051, Lubbock, TX, 79409-1051, USA
}

\date{Accepted XXX. Received YYY; in original form ZZZ}

\pubyear{2023}

\begin{document}
\label{firstpage}
\pagerange{\pageref{firstpage}--\pageref{lastpage}}
\maketitle

\begin{abstract}
Accreting low mass X-ray binaries (LMXBs) are capable of launching powerful outflows such as accretion disc winds. In disc winds, vast amounts of material can be carried away, potentially greatly impacting the binary and its environment. Previous studies have uncovered signatures of disc winds in the X-ray, optical, near-infrared, and recently even the UV band, predominantly in LMXBs with large discs ($P_{\mathrm{orb}}{\geq}20$ hrs). Here, we present the discovery of transient UV outflow features in UW CrB, a high-inclination ($i{\geq}77\degree$) neutron star LMXB with an orbital period of only $P_{\mathrm{orb}}{\approx}111~\mathrm{min}$. We present P-Cygni profiles detected for Si\,\textsc{iv} 1400{\AA} and tentatively for N\,\textsc{v} 1240{\AA} in one 15 min exposure, which is the only exposure covering orbital phase $\phi{\approx}0.7{-}0.8$, with a velocity of ${\approx}1500$~km s$^{-1}$. 
We show that due to the presence of black body emission from the neutron star surface and/or boundary layer, a thermal disc wind can be driven despite the short $P_{\mathrm{orb}}$, but explore alternative scenarios as well.
The discovery that thermal disc winds may occur in NS-LMXBs with $P_{\mathrm{orb}}$ as small as ${\approx}111$ min, and can potentially be transient on time scales as short as ${\approx}15$ min, warrants further observational and theoretical work. 

%
\end{abstract}




\section{Introduction}
\label{sec:introduction}
Accreting low mass X-ray binaries (LMXBs) are capable of producing powerful outflows.
In addition to jets, LMXBs can launch winds from the accretion disc, which can have a great impact on both the binary and its local environment.
As vast amounts of material can be carried away through disc winds \citep[e.g.][]{2002ApJ...567.1102L..Lee2002, 2012MNRAS.422L..11P..Ponti2012}, 
the accretion process can be affected \citep[e.g.][]{1983ApJ...271...70B..Begelman1983,2016Natur.534...75M..MunozDarias2016,2018Natur.554...69T..Tetarenko2018}, and the long-term orbital evolution can be altered \citep[e.g.][]{2014ApJ...784..122D..Degenaar2014, 2019A&A...627A.125M..Marino2019}. 
Outflows can also heat and stir up the surrounding interstellar medium to stimulate star formation \citep[e.g.][]{2012MNRAS.423.1641J..JusthamSchawinski2012}. 
Therefore, studying disc winds is vital in gaining a deeper understanding of X-ray binary systems and how they evolve, accretion physics, binary evolution and the role of LMXB feedback. 

Disc winds in LMXBs were initially detected using high-resolution X-ray spectroscopy through blue-shifted absorption features \citep[e.g.][]{1998ApJ...492..782U..Ueda1998,2006Natur.441..953M..Miller2006,2009Natur.458..481N..NeilsenLee2009,2012MNRAS.422L..11P..Ponti2012}. These features are proposed to result from 'hot' winds of highly ionised outflowing material. 
Disc winds have also been detected using optical and near-infrared observations, through e.g. blue-shifted absorption features and/or P-Cygni profiles.
Whereas these 'cold' winds have occasionally been detected in longer wavelengths in the past \citep[e.g.][]{1999MNRAS.306..417B..Bandyopadhyay1999}, it was not until the last few years that these started to be detected more routinely 
\citep[e.g.][]{2016Natur.534...75M..MunozDarias2016, 2020A&A...640L...3S..Sanchez-Sierras2020}.
 
Several key questions about disc winds remain, as it is currently not established exactly how these winds are launched, or how much mass is lost this way. Different mechanisms are suggested to launch disc winds in LMXBs, namely 
thermal \citep[e.g.][]{1996ApJ...461..767W..Woods1996, 2018MNRAS.479.3651H..Higginbottom2018},
radiative \citep[e.g.][]{2002ApJ...565..455P..ProgaKallman2002},
and magnetic driving \citep[e.g.][]{2000ApJ...538..684P..Proga2000, 2016A&A...589A.119C..Chakravorty2016}. However, determining the exact mechanism 
through observational studies has proven to be very challenging. 
Furthermore, the relationship between the detected hot (X-ray) and cold (optical/nIR) disc winds is not well understood yet, i.e. if these disc winds sample parts of the same outflow, or if these are resulting of two distinct outflows. Some recent studies have been performed using multi-wavelength campaigns \citep[e.g.][]{2022Natur.603...52C..CastroSegura2022,2022A&A...664A.104M..MunozDariasPonti2022} supporting a multi-phase nature of disc winds. \\
\indent A promising avenue of studying LMXB outflows is exploring features in the UV-band, as it bridges the gap between the optical/IR and X-ray band. Moreover, the UV-band contains strong line transitions of key elements like H, N, Si, O and C,  
and accretion disc spectra can peak in the UV-band as well \citep[e.g.][]{2000ApJ...539L..37H..Hynes2000}.
Outflows have been discovered using high-resolution far-UV (FUV) spectroscopy in a few studies \citep[e.g.][]{2001ApJ...562..925B..Boroson2001,2003A&A...399..211I..Ioannou2003,2010ApJ...709..251B..Bayless2010, 2022Natur.603...52C..CastroSegura2022}, but this field is still relatively unexplored. 
One aspect that greatly complicates UV-studies is interstellar extinction, especially since most LMXBs are located in the Galactic plane where the extinction along our line of sight is high \citep[see e.g.][]{2022arXiv220610053B..Bahramian2022}.\\
\indent An intriguing target for a UV outflow study is the neutron star (NS) LMXB UW CrB. The interstellar extinction to UW CrB is relatively low, on the order of $N_{\text{H}}\,{\approx}\,4\,{\times}\,10^{20}\,\mathrm{cm}^{-2}$ \citep{2016A&A...594A.116H..HI4PI} 
making it a suitable target to study in the UV. It was discovered by \cite{1990ApJ...365..686M..Morris1990} with \textit{Einstein} at a flux level of $\mathrm{f}_{\mathrm{X}}{\approx}1\,{\times}\,10^{-12}~\text{erg cm}^{-2} \text{s}^{-1}$ \citep[0.3-3.5 keV;][]{1990ApJ...356L..35G..Gioia1990}. 
One explanation for the relatively low X-ray flux of UW CrB is that the central X-ray emitting region is obscured, as it is  
a high-inclination system with $77\degree\,{\leq}\,i\,{\leq}\,81\degree$, based on the detection of eclipses both in the optical and X-ray band \citep{2005MNRAS.356.1133H...Hakala2005, 2004ApJ...608L.101H..Hynes2004, 2008ApJ...685..428M..Mason2008}. 
It is suggested to have an elliptical and precessing accretion disc, based on a measured superhump-like modulation with a period of 5.5 days, as well as the variable eclipse depth \citep[e.g.][]{2008ApJ...685..428M..Mason2008}. Moreover, the orbital period of the system is only $P_{\mathrm{orb}}{=}110.98$ minutes, indicating that UW CrB is a compact X-ray binary. 
This makes UW CrB an especially interesting target to search for disc winds, as disc winds have mostly been identified in LMXBs with an orbital period exceeding ${\gtrapprox}20$ hours \citep[see e.g.][]{2016AN....337..368D..DiazTrigo2016, 2022A&A...664A.100P..Panizo-Espinar2022}.
 
In this letter, we report on the discovery of transient outflow features in archival FUV data of UW CrB. 
%
%
%
%
%
\section{Observations, data analysis and results}
\label{sec:analysis}
\subsection{Data}
\label{sec:data}
The Hubble Space Telescope ({\it HST}) observed UW CrB on September 1, 2011 between 02:15 and 09:12 UT as part of the GTO/COS programme 12039 (PI Green).  
The data was acquired using the Cosmic Origins Spectrograph \citep[COS;][]{2012ApJ...744...60G..COS} in \texttt{TIME-TAG} mode using the primary science aperture. A total of ten exposures were obtained in five consecutive {\it HST} orbits, using two FUV (G130M and G160M) and one NUV (G230L) grating. The total exposure time is 4.7 ks for the G130M, 5.6 ks for the G160M, and 2.9 ks for the G230L gratings. This yields a wavelength coverage of $1150{-}1800$\AA{} and an average spectral resolution of $R{=}\lambda/\Delta \lambda{\approx}14000$ for the G130M and G160M gratings, and $1650{-}3200$\AA{} and $R{\approx}2650$ for the G230L grating. 
We used the {\it HST} \texttt{CalCOS} pipeline\footnote{https://github.com/spacetelescope/calcos} 
to reduce the COS data. To exclude the emission from airglow lines, the geocoronal Lyman-$\alpha$ (1208 to 1225\AA) and O\,{\sc I} (1298 to 1312\AA) line profiles were masked. 
We use the standard pipeline data products to obtain the one-dimensional spectra per exposure. 
We resampled the one-dimensional spectra on to a common wavelength grid using adapted code from the \texttt{SpectRes} package \citep{2017arXiv170505165C..SpectRes}. 
To achieve time-resolved spectroscopy, we use the \texttt{costools splittag} package to split the TIME-TAG data into sub-exposures. Light curves are extracted from the TIME-TAG events lists as described in \cite{2022Natur.603...52C..CastroSegura2022} using adapted code from the \texttt{lightcurve}\footnote{https://github.com/justincely/lightcurve} package.

\subsection{Spectral analysis} 
\label{sec:spectral_analysis}

\begin{figure}
        \centering
        \includegraphics[width=\columnwidth]{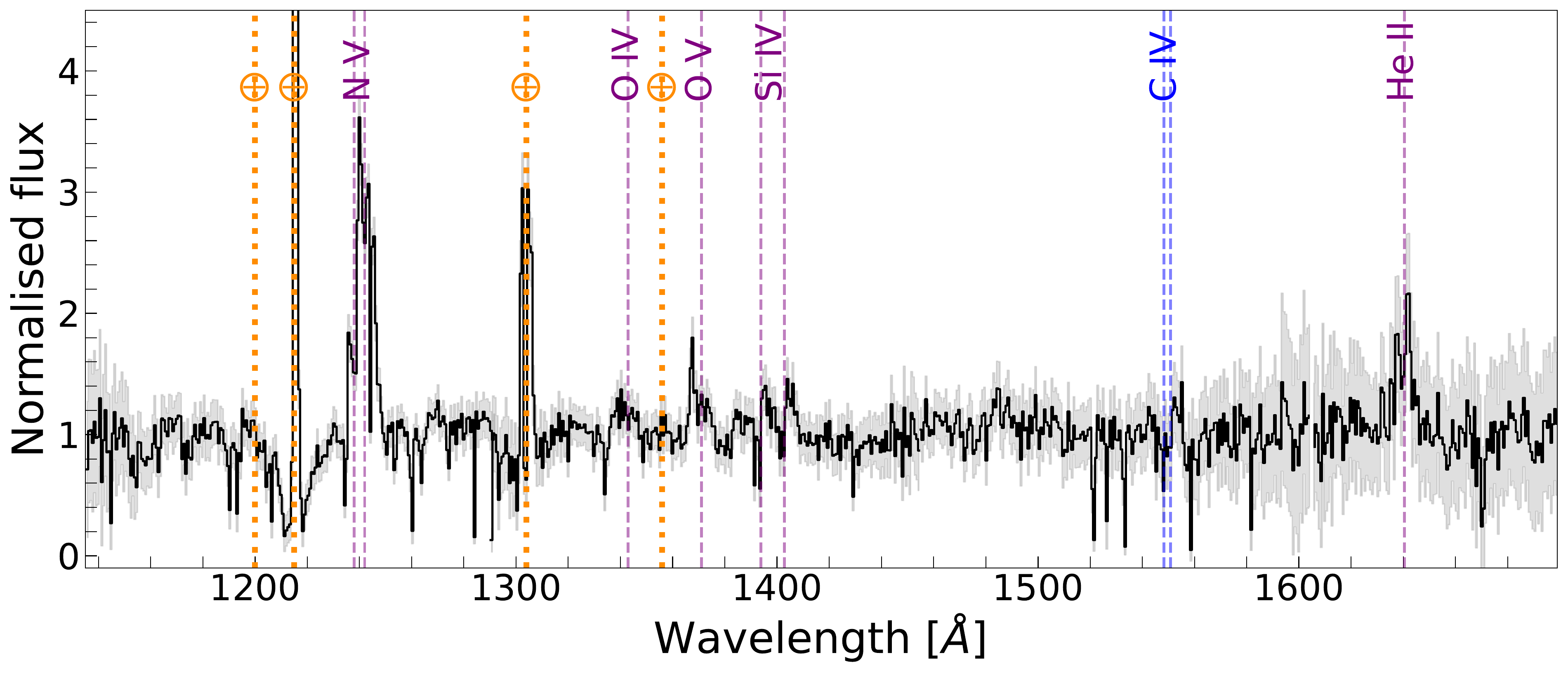}
        \caption{The normalised time-averaged FUV spectrum of UW CrB. The data are binned to 0.7{\AA} resolution, the standard errors are shown in grey. Prominent identified emission lines are labelled in purple, and other typical LMXB emission lines are labelled in blue. Terrestrial airglow emission features are marked in orange, and unlabelled narrow absorption features are interstellar.}
        \label{fig:fullspectrum}
    \end{figure}

In Figure \ref{fig:fullspectrum} we show the combined FUV spectrum.
We have not corrected the spectrum for interstellar extinction, since the estimate for the reddening along the line of sight is low, and is not expected to significantly affect the spectrum.
We also do not detect the 2175{\AA} dust feature used by other studies to perform dereddening \citep[e.g.,][]{2011ApJ...743...26F..Froning2011}, suggesting that $E(B{-}V){\lesssim}0.05$ \citep[e.g.][]{V1987A&AS...71..339V}. 

The FUV spectrum features strong emission lines, such as O\,\textsc{iv} 1343{\AA}, O\,\textsc{v} 1371{\AA}, the Si\,\textsc{iv} 1400{\AA} doublet and He\,\textsc{ii} 1640{\AA}. Most notably, it shows a strong N\,\textsc{v} 1240{\AA} doublet line, while the C\,\textsc{iv} 1549{\AA} doublet is barely identified. As these lines are both resonance lines of lithium-like ions, and as they are produced under similar physical conditions, this suggests an under-abundance of carbon in the surface layers of the donor star \citep[see e.g.][]{2002MNRAS.332..928H..Haswell2002}. This material is therefore likely to have undergone substantial CNO processing, so the donor star of this system could be an evolved main-sequence star \citep[see also][Castro Segura et al. in prep, on other LMXBs]{2011ApJ...743...26F..Froning2011, 2014ApJ...780...48F..Froning2014}.
Furthermore, no significant lines (including e.g. Mg\,\textsc{II} 2800{\AA}) are detected in the NUV spectrum. 

    \begin{figure*}
        \centering
        \includegraphics[width=\textwidth]{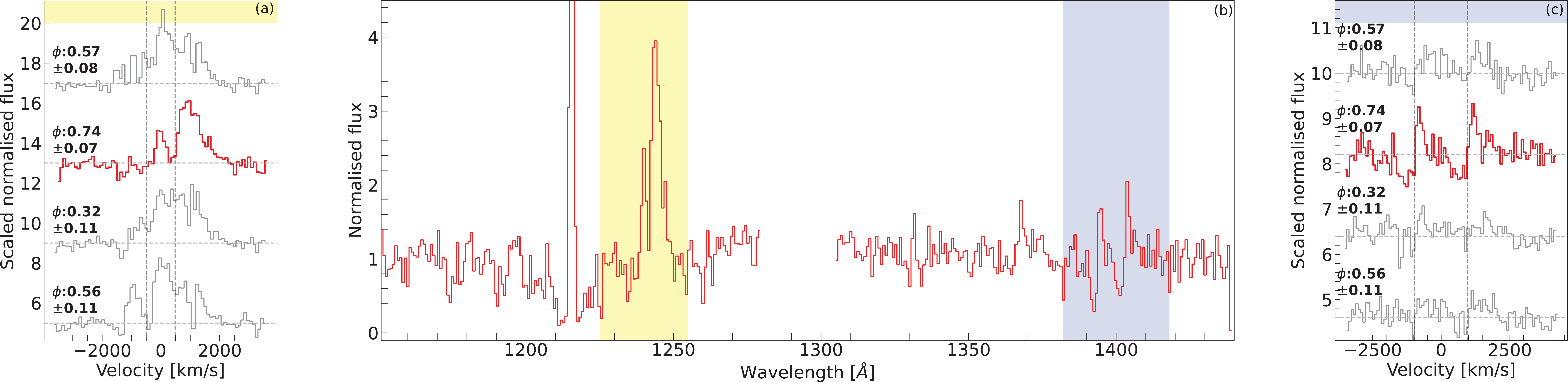}
        \caption{The outflow features detected in UW CrB. Panel (a) and (c) show the N\,\textsc{v} and Si\,\textsc{iv} line profiles in velocity space for all the available exposures, with the average orbital phase ($\phi$) per exposure indicated. Panel (b) shows the blue portion of the normalised FUV spectrum of the exposure of UW CrB showing the transient far-UV outflow features (Exposure 2) detected in Si\,\textsc{iv} 1400{\AA} and N\,\textsc{v} 1240{\AA}. The spectrum is binned to 0.8{\AA} resolution.}
        \label{fig:outflow_feature}
    \end{figure*}

When studying the spectra from the individual exposures, we find that the Si\,\textsc{iv} 1400{\AA} line profile shows clear P-Cygni profiles in one exposure. Moreover, the N\,\textsc{v} 1240{\AA} line shows tentative evidence of associated blue-shifted absorption in this same exposure. We show the spectrum of this exposure, as well as the Si\,\textsc{iv} and N\,\textsc{v} line profiles in Figure \ref{fig:outflow_feature}. 
As these lines are created at a similar temperature of around $T{\approx}10^{4}~\mathrm{K}$,  this establishes the presence of warm outflowing material in UW CrB. However, these absorption features are only detected in one of the ten total exposures (Exposure 2, specifically).
\subsection{Outflow features}
\label{sec:outflow_features}

In Exposure 2, the blue-shifted absorption troughs for Si\,\textsc{iv} and N\,\textsc{v} reach ${\approx}50$ percent below the continuum level. 
There are also very narrow absorption features, reaching to ${\approx}70$ percent below the continuum level, for both lines at around the central line-wavelengths. However, these are also seen in the other exposures, and could be attributed to interstellar absorption of the same elements. 
In Figure \ref{fig:outflow_feature} we show that the edge of the transient blue-shifted absorption features for Si\,\textsc{iv} and N\,\textsc{v} extend to around $v{\approx}{-}1500~\mathrm{km/s}$ from the central line-wavelengths, giving us the approximate terminal velocity. 

Detecting blue-shifted absorption associated with UV resonance lines, such as Si\,\textsc{iv} and N\,\textsc{v}, suggests that the outflowing material has a significant optical depth for these transitions. Based on the required minimum column densities, lower limits can placed on the mass-loss rate, as determined by \cite{2022Natur.603...52C..CastroSegura2022}.  
In this work, they rewrite the equation for the optical depth as defined in Eq. 9 of \cite{1987MNRAS.224..595D..Drew1987}, to a characteristic optical depth for C\,\textsc{iv} 1549{\AA} of: 
\begin{multline}
    \tau \simeq 33.6 \left(  \frac{f_{\mathrm{osc}}}{0.2847} \right)
    \left( \frac{\lambda}{1549{\AA}} \right)
    \left( \frac{A}{3{\times}10^{-4}} \right)
    \left( \frac{f_{\mathrm{ion}}}{1} \right)
    \left( \frac{\dot{M}_{\mathrm{wind}}}{10^{-10} \mathrm{M}_{\odot} \mathrm{yr}^{-1}} \right) \\
    \left( \frac{r}{10^{10} \mathrm{cm}} \right)^{-1}
    \left( \frac{v(r)}{1500 \mathrm{km s}^{-1}} \right)^{-2}
\end{multline}

Where $f_{\mathrm{osc}}$ and $\lambda$ are the oscillator strength and the wavelength of the line, $A$ the abundance of the relevant element relative to hydrogen, $f_{\mathrm{ion}}$ the fraction of those atoms at the correct ionisation level, $\dot{M}_{\mathrm{wind}}$ the mass-loss rate of the outflow, and $r$ the radius where velocity $v(r)$ is reached. The reference values for the velocity $v(r)$ is set on the estimated velocity of the UV outflow, and $r$ on the radius where a thermally driven wind could be launched (see Section \ref{sec:discussion}).
$f_{\mathrm{ion}}$ is set to 1 to ensure that the estimate for $\dot{M}_{\mathrm{wind}}$ is a lower limit. 
For Si\,\textsc{iv} 1400{\AA} and N\,\textsc{v} 1240{\AA}, the reference value for $A$ is based on solar abundances obtained from \cite{2009LanB...4B..712L..Lodders2009}, which are $3.41{\times}10^{-5}$ and $7.24{\times}10^{-5}$ for these lines, respectively. The reference value for $f_{\mathrm{osc}}$ is adopted from \cite{2003ApJS..149..205M..Morton2003}, which are $0.767$ and $0.234$ for the lines, respectively. With the assumption that $\tau{\gtrapprox}1$ based on the depth of the lines, the estimated lower limits on the mass-loss rates are $\dot{M}_{\mathrm{wind}}{\gtrapprox}\,1.11{\times}10^{-11}\,\mathrm{M}_{\odot} \mathrm{yr}^{-1}$ for Si\,\textsc{iv}, and $\dot{M}_{\mathrm{wind}}{\gtrapprox}\,1.89{\times}10^{-11}\,\mathrm{M}_{\odot} \mathrm{yr}^{-1}$ for N\,\textsc{v}.
This calculation does assume a symmetric persistent outflow, see Section \ref{sec:light_curve} and \ref{sec:discussion} for discussion on the nature of the detected outflow. 
\subsection{Light curve}
\label{sec:light_curve}

To study the time-dependent properties of the outflow, we study the extracted FUV light curve.
\cite{2012AJ....144..108M..Mason2012} calculated the eclipse ephemeris of UW CrB using a total of 56 eclipses spanning 20 years, allowing us to determine the orbital phase of the {\it HST} observations.

\begin{figure}
        \centering
        \includegraphics[width=\columnwidth]{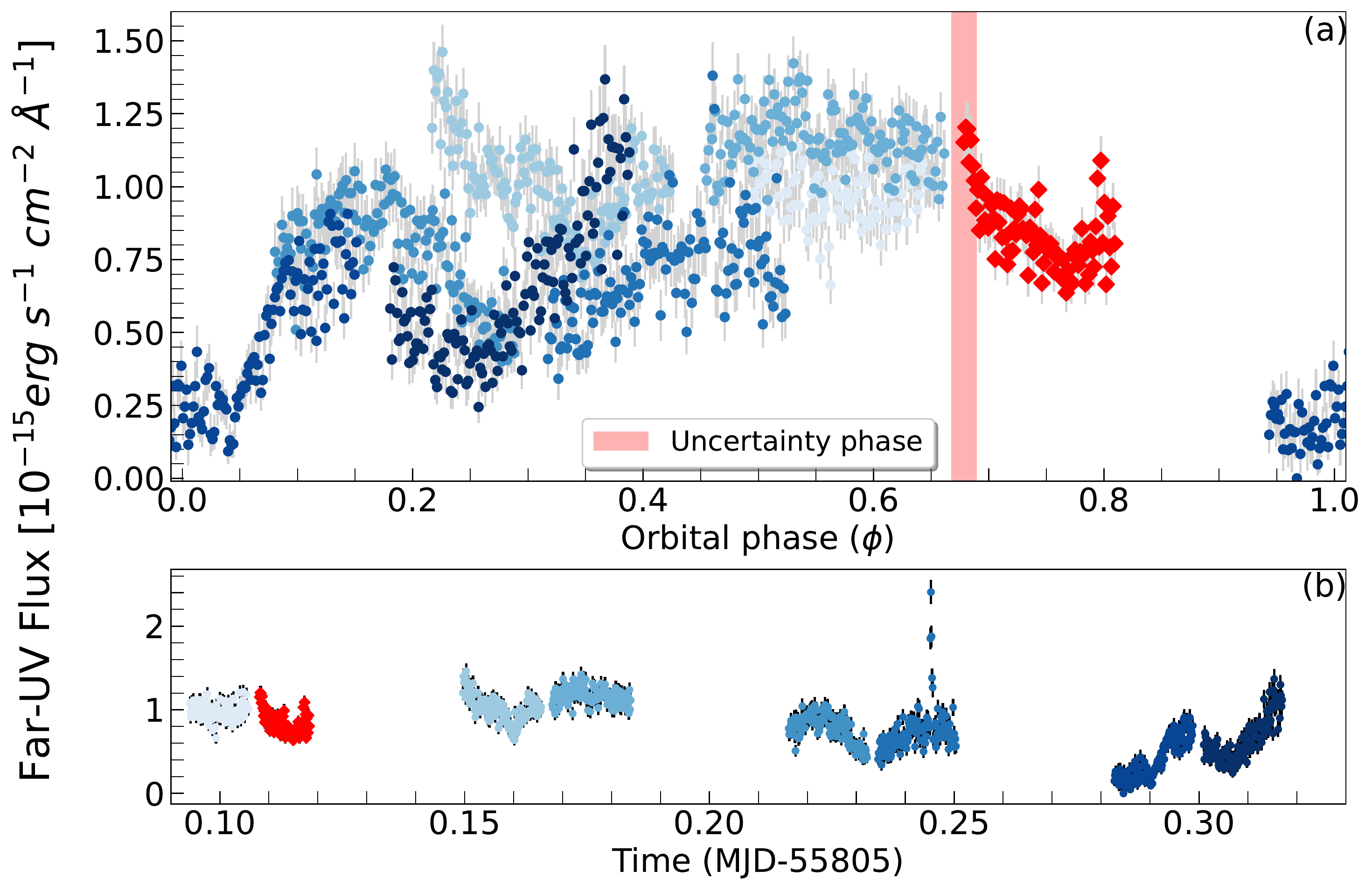}
        \caption{The light curve for all the FUV {\it HST} COS exposures. The exposure where the outflow features are detected (Exposure 2) is shown in red, and other exposures are indicated with blue sequential colours from light to dark to show the evolution in time. 
        A type-I burst is identified around MJD 55805.23, but is cropped from panel (a) for clarification.
        Upper panel (a) shows the phase-folded light curve, where orbital phase $\phi{=}0$ indicates the mid-point of the eclipse. The uncertainty on the orbital phase is indicated with a red bar.
        Lower panel (b) shows the light curve plotted with time on the x-axis in days.}
        \label{fig:LC}
    \end{figure}

We show the light curve obtained from the FUV {\it HST} COS data in Figure \ref{fig:LC}, with the phase-folded light curve shown in the upper panel (a), and the full light curve shown in lower panel (b).
From the phase-folded light curve we find that the lowest UV flux is found at $\phi{\approx}0$, which is expected to be the center of the eclipse. The lowest flux is not exactly at $\phi{=}0$, which could be due to the orbital phase of mid-eclipse wandering by up to $\Delta\phi \pm 0.08$ \citep{2012AJ....144..108M..Mason2012} and the uncertainty on the orbital phase being around $\delta{\phi}{\approx}0.011$. 

We find that the exposure where we detect the outflow features (Exposure 2), is the only exposure covering the orbital phase range $\phi{\approx}0.7{-}0.8$. Therefore, it could be that the outflow is dependent on the orbital phase. 
Other parts of the orbital phase range are covered by one or more exposures, but no outflow features are detected here. 
Alternatively, it could be a transient outflow. Optical P-Cygni profiles have been observed during flaring activity of some BH-LMXBs \citep[e.g.][]{2016Natur.534...75M..MunozDarias2016, 2022A&A...664A.104M..MunozDariasPonti2022}.
However, we do not identify any flaring components in the UV light curve.  

Lastly, the outflow features could also appear transient due to line-of-sight effects. Based on complex variations in the optical and X-ray light curves, \cite{2008ApJ...685..428M..Mason2008} and \cite{2009MNRAS.394..892H..Hakala2009} suggest that UW CrB has a precessing elliptical accretion disc. \cite{2012AJ....144..108M..Mason2012} also discuss that variations in the optical light curve could be caused by an out-of-plane structure such as a non-axisymmetric flared or warped accretion disc, or a non-axisymmetric disc wind \citep[as observed for X1822-371; see e.g.][]{2010ApJ...709..251B..Bayless2010}. This could be causing the large-amplitude variations in the UV light curve. 
The outflowing material could therefore potentially be obscured from the line-of-sight, resulting in the \mbox{transient UV outflow features}. 

\subsection{Time-resolved spectroscopy}
\label{sec:time_resolved_spectroscopy}

\begin{figure}
        \centering
        \includegraphics[width=\columnwidth]{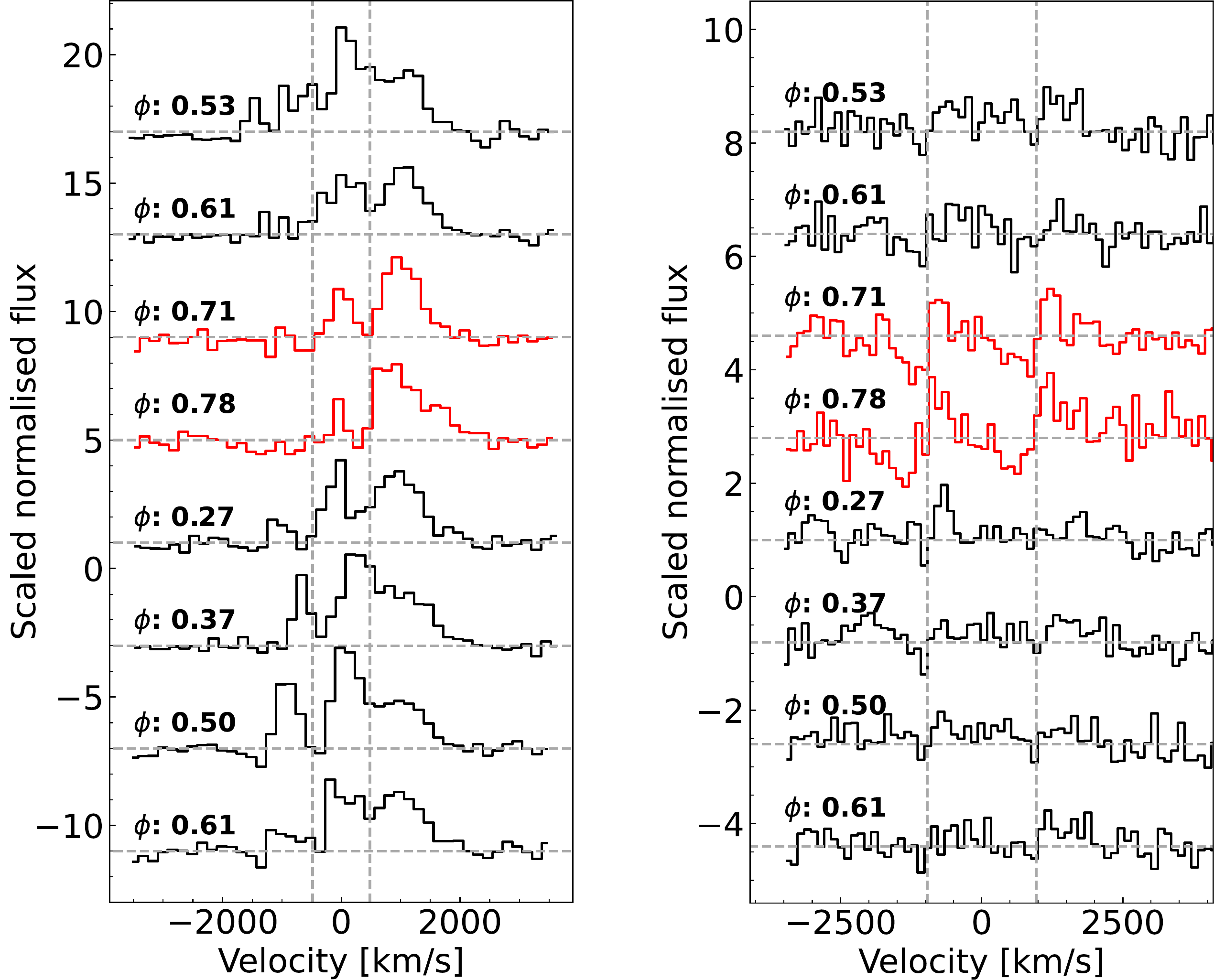}
        \caption{The N\,\textsc{v} 1240{\AA} (right panel) and Si\,\textsc{iv} 1400{\AA} (left panel) lines in the time-resolved spectra of sub-exposures of UW CrB. The average orbital phase of the sub-exposures are indicated on the upper-left, within a range of around $\Delta\phi\,{\pm}\,0.04$. 
        The sub-exposures of Exposure 2 are indicated in red. The other sub-exposures are indicated in black.
        }
        \label{fig:time-resolved}
    \end{figure}

To explore the time-dependent properties of the outflow in more detail, we split each original exposure in two, to have enough signal in each sub-exposure. 
In Figure \ref{fig:time-resolved} we show sub-exposures of the Si\,\textsc{iv} and N\,\textsc{v} lines. 
We find that the P-Cygni profiles found for Si\,\textsc{iv} and N\,\textsc{v} do not appear to change significantly in time during Exposure 2, as shown for the two sub-exposures indicated in red. 
In all other sub-exposures, we find no evidence of P-Cygni profiles for any of the lines in the spectrum.
There appear to be hints of associated blue-shifted absorption features for Si\,\textsc{iv} in the sub-exposures preceeding Exposure 2. However, we do not see this for the N\,\textsc{v} line. 
    
The line profile of N\,\textsc{v} appears to change during the \mbox{(sub-)exposures}. The line appears double peaked during some \mbox{(sub-)exposures}, and single peaked during others. Moreover, the line centres of the two identified peaks for N\,\textsc{v} also appear to shift, between around ${\approx}{-}500$ to $500$ km/s. However, it is not clear if these are the two peaks of the doublet, or if the narrow left-most peak is a separate feature. Furthermore, it is unclear if the shift in the line-centres is related to a doppler-shift, or if this is related to other changes in the line-profile, especially since it appears differently for sub-exposures covering similar parts of the orbital phase. 
%
%
\vspace{-0.3cm}
\section{Discussion}
\label{sec:discussion}

Based on our detection of the transient UV outflow features for UW CrB, we explored if outflow features are detected in archival studies using other wavebands. 
\cite{1990ApJ...365..686M..Morris1990} present optical spectra of UW CrB, and note that the He\,{\sc I} lines are systematically blue-shifted relative to the emission lines, potentially indicating a cool outflow. 
\cite{2005MNRAS.356.1133H...Hakala2005} present 0.2-10 keV X-ray spectra. 
NS-LMXBs typically show broad Fe-K emission lines, interpreted as reflection from the inner disc \citep[e.g.][]{2008ApJ...674..415C..Cackett2008}. 
However, the emission line in the X-ray spectrum of UW CrB appears narrow. While not conclusive, this may indicate that it results from scattering in a wind instead of disc reflection \citep[e.g.][]{2009ApJ...700.1831T..Titarchuk2009}. 

We can consider what types of outflows would be possible in this system, starting with the scenario of disc winds. The estimated velocity of 1500~km s$^{-1}$ is consistent with the outflow speeds detected for (thermally-driven) disc winds \citep[see overviews in e.g.][for other wavebands]{2016AN....337..368D..DiazTrigo2016,2022A&A...664A.100P..Panizo-Espinar2022}. 
Thermally driven winds can be launched from 10 percent of the Compton radius \citep{1983ApJ...271...70B..Begelman1983}, which is the radius at which the local isothermal sound speed (at the Compton temperature, $T_{\mathrm{IC}}$) is equal to the escape velocity in the disc. It is defined as $R_{\mathrm{IC}}{=}\frac{G M\mu m_{\mathrm{H}}}{k_{\mathrm{B}} T_{\mathrm{IC}}}$, where $M$ is the mass of the central object and $\mu$ is the mean molecular mass. We assume $M{=}1.4\mathrm{M}_{\odot}$ for a neutron star, and that $\mu{=}0.6$ and $T_{\text{IC}}{=}1.4\times10^{7}$ K as in \cite{2017ApJ...836...42H..Higginbottom2017}, resulting in \mbox{$R_{\text{IC}}{\approx}9\times10^{10}~\mathrm{cm}$} for UW CrB.
To estimate the size of the disc, we can use two methods. 
We can calculate the circularisation radius, which is the radius within which the accretion disc is formed from the accretion flow,
or the 3:1 resonance radius, which is the radius at which the disc is expected to be truncated for systems with low eccentricity. 
For both calculations, we use the mass-ratio $q{\approx}0.15$, the mass $M_{1}{\approx}1.4\,\mathrm{M}_{\odot}$ for the compact object, and the mass $M_{2}{\approx}0.2\,\mathrm{M}_{\odot}$ for the companion star, as proposed by \cite{2008ApJ...685..428M..Mason2008}. We find both the circularisation and 3:1 resonance radius are around $R{\approx}4\times10^{10}~\mathrm{cm}$, thus the disc likely exceeds $0.1 R_{\mathrm{IC}}$ needed to launch a thermal wind.

To explore this a bit further, we have modelled the expected thermal stability curves for UW CrB, following the procedure of \cite{2017ApJ...836...42H..Higginbottom2017}. Briefly, we use the radiative transfer and ionization code {\sc python} \citep{LK2002ApJ...579..725L, 2018MNRAS.479.3651H..Higginbottom2018, H2019MNRAS.484.4635H, H2020MNRAS.492.5271H} to determine the physical conditions in an irradiated, optically thin parcel of gas. The conditions in such a parcel are controlled by two factors: (i) the spectral energy distribution of the radiation field, and (ii) the strength of the irradiation. 
As the central X-ray source of UW CrB is likely obscured, we do not have access to the intrinsic spectral energy distribution (SED) of the system. We therefore use the simple, but flexible analytical model $F_{\nu}{=}\nu^{-0.2} \text{exp}(-h\nu/k_{\text{B}}T_{\text{X}})$ to describe the SED. 
This form was proposed for the NS-LMXBs 4U 0614+091 in \cite{2010ApJ...710..117M..Migliari2010}, which has $P_{\mathrm{orb}}$ and $L_{\mathrm{X}}$ comparable to UW CrB. 
Out of the few NS-LMXBs with reported SEDs, the SED of 4U 0614+091 seems most representative for UW CrB, despite the factor ${\approx}2$ difference in $P_{\mathrm{orb}}$.
We then consider two different possibilities: (i) $T_{\text{X}}{\approx}4{\times}10^{6}~\mathrm{K}$, representating a disc black body SED, and (ii) $T_{\text{X}}{\approx}1.3{\times}10^{7}~\mathrm{K}$, representating a central black body SED. The strength of the irradiation is controlled by the ionization parameter $\xi{=} 
L_{\text{X}} / n R^2$, where $L_{\text{X}}$ is the source luminosity, $n$ is the gas density and $R$ is the distance of the gas parcel from the source. 

\begin{figure}
        \centering
        \includegraphics[width=\columnwidth]{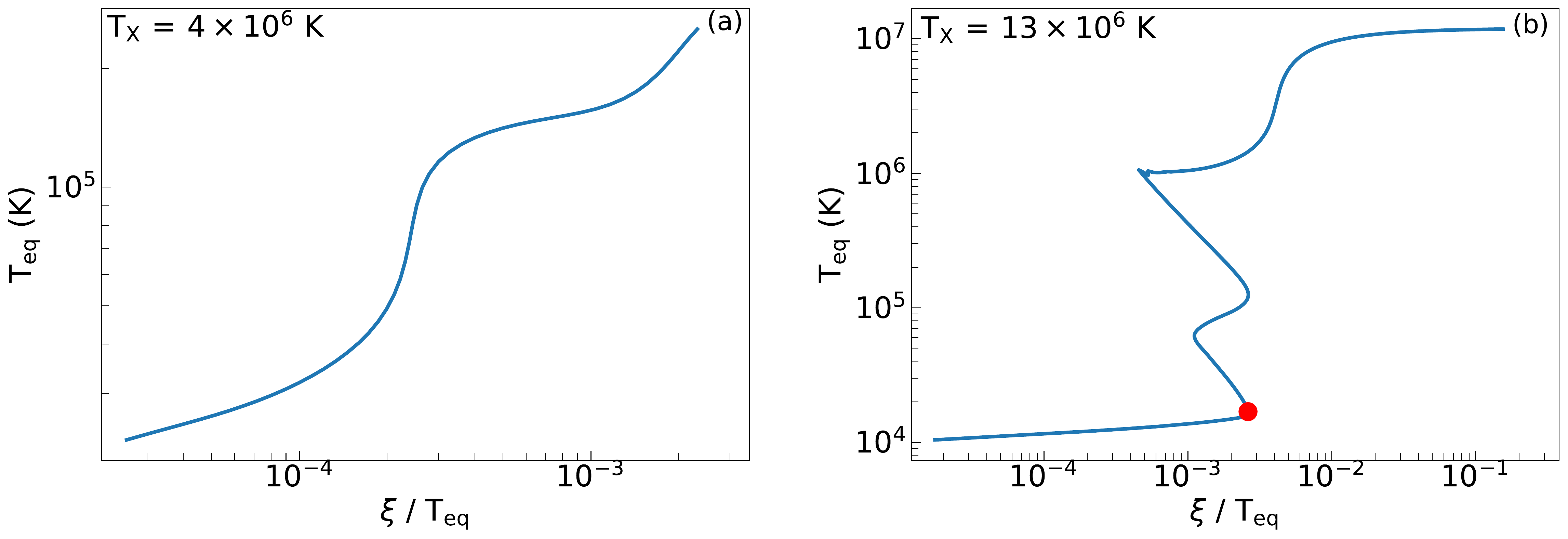}
        \caption{Thermal stability curves modelled for UW CrB, using a disc black body (Panel a) and a central black body (Panel b). The red dot in \mbox{Panel b} indicates the critical ionization parameter; $\xi_{\text{cool,max}}\,{=}\,44.2$.}
        \label{fig:thermal-stability-curves}
    \end{figure}

As shown in Fig. \ref{fig:thermal-stability-curves}, the stability curve is then a plot of the equilibrium temperature, $T_{\mathrm{eq}}$, against $\xi/T_{\mathrm{eq}}$ (which traces the ratio of radiation to gas pressure). Moving from left to right on this plot corresponds to moving vertically upwards in the irradiated atmosphere. The gas is thermally stable (unstable) if the slope of $T_{\mathrm{eq}}(\xi/T_{\mathrm{eq}})$ is positive (negative). In order to launch a powerful irradiation-driven outflow, the material in the disc atmosphere needs to experience rapid runaway heating, i.e. it needs to become  thermally unstable at some critical height in the atmosphere. Fig. \ref{fig:thermal-stability-curves} shows that this condition is not met for the cooler 'disc' SED, which is thermally stable everywhere. However, the hotter 'central black body' SED, representing emission from the NS surface and/or boundary layer, becomes thermally unstable at a critical ionization parameter $\xi_{\text{cool,max}}\,{=}\,44.2$. In this case, material in the atmosphere would experience explosive heating, and a strong thermally-driven wind is likely to be launched. 
 
We also explored the possibility of other wind driving mechanisms, i.e. radiative and magnetic driving.
Radiation pressure due to electron scattering can assist thermal expansion to drive a disc wind if the luminosity is within around a factor of ${\approx}2$ of the Eddington limit \citep[see e.g.][]{2002ApJ...565..455P..ProgaKallman2002}. 
Based on the known $P_{\mathrm{orb}}$ and measured optical brightness, we can use the relation of \cite{1994A&A...290..133V..vanParadijs1994} to estimate the intrinsic $L_{\mathrm{X}}$, which suggests it is on the order of ${\approx}0.1L_{\mathrm{EDD}}$. Such an accretion luminosity would be consistent with the amplitude, recurrence rate and duration of the type-I bursts reported by e.g. \cite{2005MNRAS.356.1133H...Hakala2005}.
Therefore, we conclude that the outflow is unlikely to be caused by a radiatively driven wind in UW CrB. 
On the other hand, magnetic field lines threading the disc can launch a wind anywhere in the disc \citep[see e.g.][]{2016A&A...589A.119C..Chakravorty2016}.
Based on our data, we were unable to confirm or exclude a magnetically driven disc wind.   

Finally, we briefly consider a few other possible scenarios for the outflow in UW CrB. 
Firstly, the P-Cygni profiles could alternatively have formed in a wind of matter evaporating from the atmosphere of the companion star. 
Some NS-LMXBs such as EXO 0748-676 are proposed to ablate material from the companion star, through a pulsar wind or X-ray heating \citep[see e.g.][]{2012MNRAS.420...75R..Ratti2012, 2022MNRAS.510.4736K..Knight2022}. 
Secondly, we note that the estimated velocity of the UV outflow of ${\approx}1500$ km s$^{-1}$ is around the Keplerian velocity of the outer disc in UW CrB \citep[see e.g.][]{1987A&A...178..137F..Frank1987}. So potentially, the detected outflow could result from the accretion stream and/or the hot spot where the accretion stream impacts the disc. This hot spot is expected to be visible at orbital phase $\phi{\approx}0.6{-}0.8$ \citep{1987A&A...178..137F..Frank1987, 2002A&A...382..130I..Ioannou2002} which matches the orbital phase where the UV outflow is detected. However, little is known about such outflows. 
Follow-up observations are needed to study possible phase-dependence of the outflow, and to confirm outflow features in optical and X-ray.

\vspace{-0.3cm}
\section*{Acknowledgements}
We thank M. Middleton, J. Matthews, M. Diaz Trigo and K. Long for useful discussions, and M. Stoop for helpful comments. 
SF and ND acknowledge the hospitality of the University of Southampton, where part of this research was carried out. This research is based on observations made with the NASA/ESA Hubble Space Telescope. These observations are associated with program(s) 12039.
%
\vspace{-0.3cm}
\section*{Data Availability}

The data underlying this article will be available in Zenodo at DOI: 10.5281/zenodo.7883747 upon publication.
This astrophysical data set was retrieved from archival UV data from sources in the public domain: \url{https://mast.stsci.edu/search/ui/#/hst}. 


\vspace{-0.3cm}






\bsp	
\label{lastpage}
\end{document}